\documentclass[%
 aip,
 amsmath,amssymb,
 reprint,%
]{revtex4-1}

\usepackage{graphicx}
\usepackage{dcolumn}
\usepackage{bm}

\usepackage[utf8]{inputenc}
\usepackage[T1]{fontenc}
\usepackage{mathptmx}
\usepackage{siunitx}
\usepackage{textcomp}
\usepackage{xcolor}
\usepackage{float}

\begin{document}

\preprint{AIP/123-QED}

\title[Photo-assisted negative ion production]{Experimental evidence on photo-assisted O$^-$ ion production from Al$_2$O$_3$ cathode in cesium sputter negative ion source}

\author{O. Tarvainen}%
\email{olli.tarvainen@stfc.ac.uk}
\affiliation{STFC ISIS Pulsed Spallation Neutron and Muon Facility, Rutherford Appleton Laboratory, Harwell, OX11 0QX, UK\looseness=-1}

\author{R. Kronholm}
\affiliation{University of Jyv\"{a}skyl\"{a}, 40500 Jyv\"{a}skyl\"{a}, Finland}

\author{M. Laitinen}
\affiliation{University of Jyv\"{a}skyl\"{a}, 40500 Jyv\"{a}skyl\"{a}, Finland}

\author{M. Reponen}
\affiliation{University of Jyv\"{a}skyl\"{a}, 40500 Jyv\"{a}skyl\"{a}, Finland}

\author{J. Julin}
\affiliation{University of Jyv\"{a}skyl\"{a}, 40500 Jyv\"{a}skyl\"{a}, Finland}

\author{V. Toivanen}
\affiliation{University of Jyv\"{a}skyl\"{a}, 40500 Jyv\"{a}skyl\"{a}, Finland}

\author{M. Napari}
\affiliation{University of Southampton, Southampton SO17 1BJ, UK}

\author{M. Marttinen}
\affiliation{University of Jyv\"{a}skyl\"{a}, 40500 Jyv\"{a}skyl\"{a}, Finland}

\author{D. Faircloth}
\affiliation{STFC ISIS Pulsed Spallation Neutron and Muon Facility, Rutherford Appleton Laboratory, Harwell, OX11 0QX, UK\looseness=-1}

\author{H. Koivisto}
\affiliation{University of Jyv\"{a}skyl\"{a}, 40500 Jyv\"{a}skyl\"{a}, Finland}

\author{T. Sajavaara}
\affiliation{University of Jyv\"{a}skyl\"{a}, 40500 Jyv\"{a}skyl\"{a}, Finland}

\date{\today}

\begin{abstract}

The production of negative ions in cesium sputter ion sources is generally considered to be a pure surface process. It has been recently proposed that ion pair production could explain the higher-than-expected beam currents extracted from these ion sources, therefore opening the door for laser-assisted enhancement of the negative ion yield. We have tested this hypothesis by measuring the effect of various pulsed diode lasers on the O$^-$ beam current produced from Al$_2$O$_3$ cathode of a cesium sputter ion source. It is expected that the ion pair production of O$^-$ requires populating the 5d electronic states of neutral cesium, thus implying that the process should be provoked only with specific wavelengths. Our experimental results provide evidence for the existence of a wavelength-dependent photo-assisted effect but cast doubt on its alleged resonant nature as the prompt enhancement of beam current can be observed with laser wavelengths exceeding a threshold photon energy. The beam current transients observed during the laser pulses suggest that the magnitude and longevity of the beam current enhancement depends on the cesium balance on the cathode surface. We conclude that the photo-assisted negative ion production could be of practical importance as it can more than double the extracted beam current under certain operational settings of the ion source.
\end{abstract}

\maketitle

\section{\label{introduction}Introduction}

\begin{figure*}[!htb]
\centering
\includegraphics[width=\textwidth]{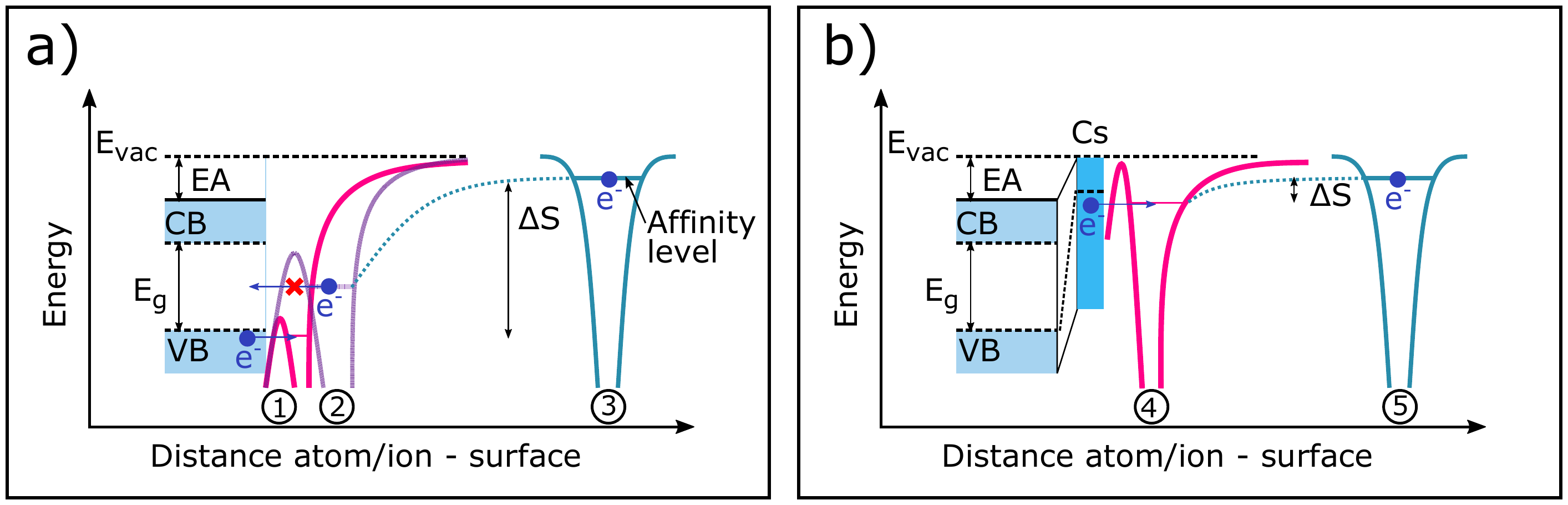}
\caption{\label{surfaceionization}A sketch of the electron capture process of atoms released by heavy ion bombardment from (a) insulator and (b) cesium covered insulator. The numbers indicate selected positions of the outbound atom/ion relative to the material surface depicted by the quantum mechanical potential well and explained in the bulk text.}
\end{figure*}

The negative ion production in cesium sputter ion sources is traditionally attributed to surface ionization by resonant tunneling of an electron from the conduction band or interstitial sites of a metal or compound material covered with a layer of cesium to the affinity level of a neutral atom and subsequent ejection of the negative ion from the surface. A general review summarizing the relevant steps of the process can be found in the literature\cite{Alton} (and original references therein). The ionization efficiency depends strongly on the work function of the surface as first noted by Yu\cite{Yu}, the sputtering yield and the escape velocity of the negative ions \cite{vanOs}. Thus, cesium atoms and ions play a dual role in the process; the alkali metal coverage reduces the surface work function while the heavy ion bombardment releases negative ions from the surface at sufficient energy to overcome the force exerted by the induced image charge within the material causing detachment of the anions\cite{Rasser}.

The surface production of negative ions by resonant tunneling is unquestionable as witnessed also in plasma ion sources serving as H$^{-}$/D$^{-}$ injectors for large-scale accelerator facilities \cite{Faircloth} and neutral beam injectors for thermonuclear fusion \cite{Hemsworth}. However, it is plausible that there are other mechanisms contributing to the negative ion yield. This claim is supported by the discrepancy between the yields of negative ions deduced with reasonable estimates of the electron affinities and work functions involved, and the measured negative ion currents from cesium sputter ion sources, first acknowledged by Krohn \cite{Krohn}. Despite of the obvious disagreement between the theory and observations, there are no systematic experiments attempting to solve the mystery, which is probably due to the lack of a testable hypothesis on the mechanism governing the negative ion production through an alternative path. Thus, the majority of the notions in the literature are anecdotal, a famous example being the assertion of the strong belief of Middleton that "the ionization occurs primarily in the blue-glowing plasma of Cs created in sputter-induced pits or in purposefully recessed samples" \cite{Middleton}. The glow is presumably sustained by electron impact excitation of neutral Cs to the 7p states by secondary electrons emitted from the cathode and the subsequent de-excitation back to the 6s ground state emitting blue light. 

This work has been inspired by a recent publication by Vogel \cite{Vogel}, which introduces a physical mechanism that would not only explain the enhanced yield of negative ions but can also be probed in a controllable experiment. This hypothesis is based on resonant ion-pair production first noted by Lee et al. in thermal alkali vapors \cite{Lee}. The ion pair-production is described by the (chemical) reaction equation A$^*+$B $\rightarrow$ A$^{+}+$B$^{-}$, which depicts the interaction between a neutral atom on an excited state (A$^{*}$) and a ground state neutral atom (B) with positive electron affinity resulting to the formation of positive (A$^{+}$) and negative (B$^{-}$) ion pair. According to so-called Landau-Zener-Stückelberg (LZS) formalism the probability, i.e. cross section, of the ion-pair production depends strongly on the energy difference of the electron donor ionization potential from the excited state and the electron affinity of the electron acceptor \cite{Grice, Buslov, Narits}. In practice this means that excitations to specific electronic states of the donor atoms can enhance the negative ion production of those anions with matching electron affinities (see Section~\ref{pairproduction} for an example illustration). These excitations can occur as a result of inelastic collisions between neutral Cs atoms and electrons emitted from the cathode or they can be facilitated externally by photon absorption.   

The above-said paper describes a proof-of-concept experiment where the C$^-$ beam current extracted from a cesium sputter ion source was enhanced by approximately 10\% by shining a \SI{450}{\nano\meter} / \SI{5}{\watt} laser beam radially into the recessed cathode \cite{Vogel}. It is argued that approximately \SI{20}{\micro\watt} of the power was absorbed by the excitation of the Cs(7p)-states, which are said to be in resonance with the effective ionization potential $I_{p,\textrm{eff}}=E_A+\Delta E$ in the reaction Cs$(7p)+$C $\rightarrow$ Cs$^{+}+$C$^{-}$. Here $E_A$ is the electron affinity of the anion and $\Delta E$ the endothermicity of the reaction. A direct quote from the paper \cite{Vogel} reads: "The quiescent \SI{40}{\micro\ampere} current of C$^-$ immediately jumped to \SI{45}{\micro\ampere} when the laser passed across the front of the sample within the 3 sec time resolution of data collection.  The source held the higher current as long as the laser passed behind the immersion lens, but no long term data was taken." 

Given the temporal resolution of the experiment, lack of actual data and staggering nearly 70\% efficiency of the pair-production (see section \ref{discussion} for further discussion) reported in the original publication \cite{Vogel}, we decided to subject the alleged mechanism to further scrutiny attempting to compare it to alternative explanations of enchanced negative ion production. All experimental data described hereafter was obtained with O$^-$ ion beams extracted from an Al$_2$O$_3$ cathode of a cesium sputter ion source. The choice of O$^-$ and its pair production mechanism in interaction with electronically excited Cs atoms are explained in Section \ref{pairproduction}. The experimental setup including the SNICS ion source, the adjacent beamline and the diode lasers employed for the experiment are described in Section \ref{setup}. Finally, the experimental results and conclusions are presented in Sections \ref{results} and \ref{discussion}.

\section{\label{theory}Negative ion formation}

Interpretation of the experimental data requires understanding the principles of negative ion surface production via cesium sputtering as well as the possible Cs(5d)~+~O~$\rightarrow$~Cs$^+$~+~O$^-$ ion pair production mechanism. A qualitative description of each relevant process is therefore given below. These include electron transfer to the electron affinity level of the anion either directly from the material or via photoelectron emission, bond-breaking of chemical compounds and the possible ion pair production mechanism.

\begin{figure*}[!htb]
\centering
\includegraphics[width=\textwidth]{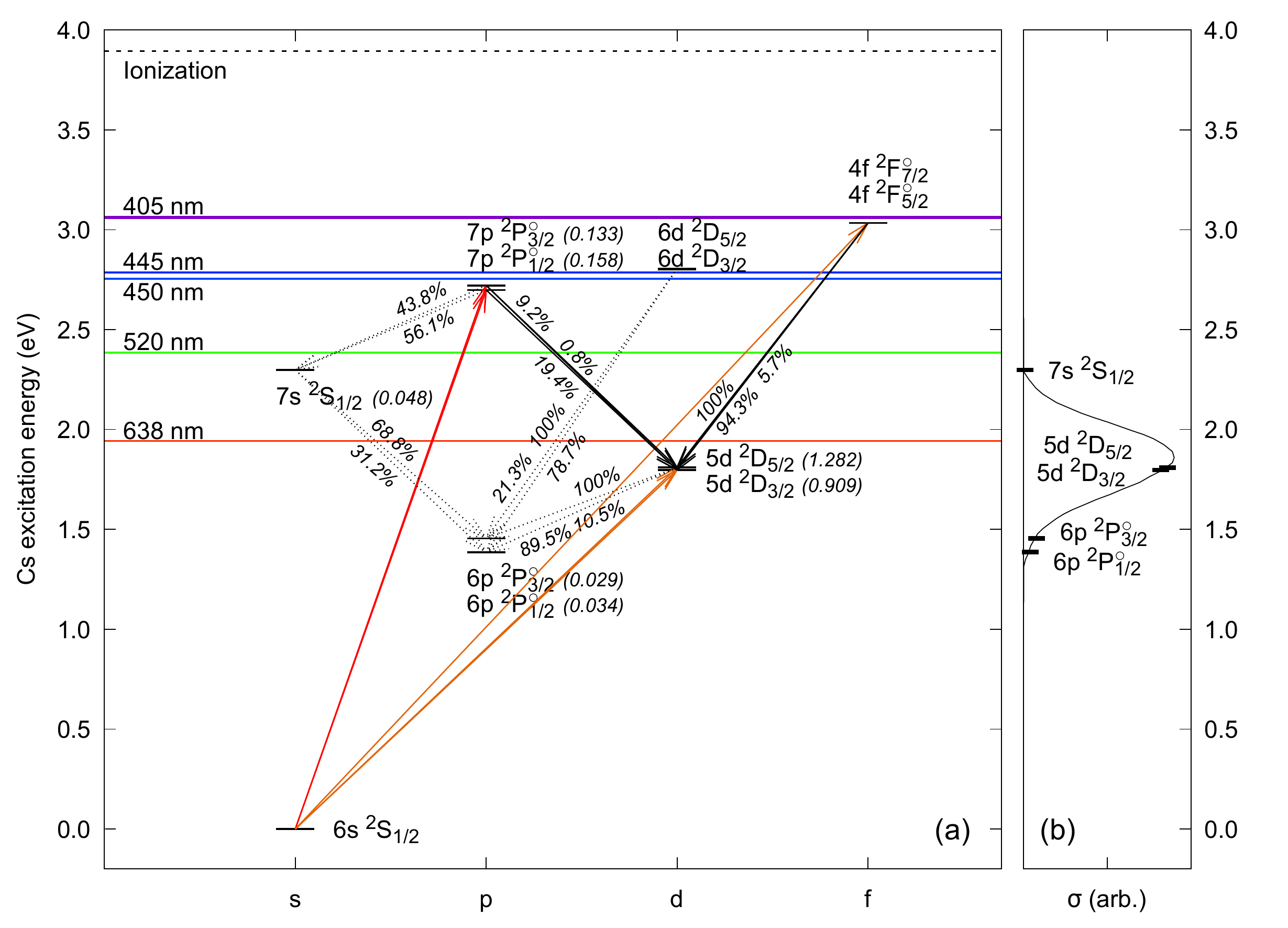}
\caption{\label{grotrian}A partial Grotrian diagram of neutral cesium. The diagram (a) shows the putative excitations corresponding to the neutral cesium electronic energy levels \cite{NIST} and the wavelengths of the diode lasers used in our experiment. The red arrows indicate the optically allowed transitions to 7p states and the orange arrows the energetically possible but very unlikely transitions to 5d and 4f. The photon energies of each laser are marked with horizontal lines with the line colors matching with the corresponding wavelenghts and the line widths to the FWHMs. The radiative decays populating the 7s, 5d and 6p -states that overlap in energy with the normalized O$^-$ ion pair production cross section \cite{Vogel} $\sigma$ (b) are indicated by black arrows. The most likely excited state contributing to the O$^-$ pair production is the metastable 5d (populated by the transitions marked with solid downward arrows) almost matching the peak of the cross section as indicated by the projected black bars corresponding to the fine structure of the electronic state(s). The spontaneous lifetime of each excited state in \textmu s is shown in parentheses. The branching ratios of the de-excitations calculated from the reported oscillator strengths\cite{Warner} are displayed next to each downward transitions. Two branching ratios associated with a single arrow correspond to the transition to/from different fine structure levels of the excited states.}
\end{figure*}

\subsection{\label{surfaceproduction}Surface production of negative ions by cesium sputtering}

The negative ion production by cesium sputtering is described schematically in Fig.~\ref{surfaceionization} for (a) bare insulating material (e.g. Al$_2$O$_3$) and (b) Cs covered insulating material. The material surface is exposed to bombardment by Cs$^+$ ions ejecting atoms from the bulk material. The electron affinity level of these atoms depends on the distance from the material surface and the material properties affecting the Debye screening radius \cite{Rasser}. At close proximity of the surface, denoted with (1) in Fig.~\ref{surfaceionization}, the affinity level of the outbound atom overlaps (in energy) with the valence band (VB) of the insulating material allowing electron tunneling through the potential barrier and formation of the anion. As the anion moves away from the surface (2), the reverse process i.e. electron tunneling back into the material is prohibited as the affinity level now overlaps with the band gap ($E_g$) between the valence and conduction (CB) bands. Far from the surface (3) the electron affinity level reaches the value of a free atom (ion), e.g. \SI{1.46}{\electronvolt} below the vacuum level for oxygen. The negative ion yield of the process can be greatly enhanced by deposition of Cs atoms onto the material surface as depicted in Fig.~\ref{surfaceionization}(b). This is because Cs modifies the band structure lowering the effective work function of the surface. In this case (4) the quantum-mechanically allowed interaction distance of electron tunneling to the atoms sputtered from the bulk material is much longer. That is because the required downshift of the electron affinity $\Delta S$ resulting in overlap with electronic states of the surface layer is reduced in comparison to bare insulator. For both cases shown in the figure $\Delta S$ is referenced to the value of the electron affinity far from the surface (5). Detailed accounts and surface material / ion-specific variations of the described process for metals and insulators can be found from the literature \cite{Alton, Rasser, Cartry}.

For the purpose of this paper it is important to acknowledge that negative ions can be produced directly from compound materials, e.g. Al$_2$O$_3$, by heavy ion bombardment induced bond-breaking of the molecular solids into cation-anion pairs \cite{Williams, Slodzian}. The relative importance of such direct negative ion production and the electron capture by neutral sputtered particles as explained above is unknown as the probability of each process depends on the material, Cs coverage and cathode bias. Finally, it could be argued that photon absorption might increase the negative ion yield by allowing photoelectrons to overcome the surface potential barrier and become bound to the affinity level of the ejected atom, thus supplementing the described electron capture by tunneling. To our knowledge the possible role of photon absorption in the electron capture process has not hitherto been studied whereas Blahins et al. have recently used a laser to study photodetachment with a cesium sputter source \cite{Blahins}. 

\subsection{\label{pairproduction}Resonant ion pair production}

The ion pair production cross section depends on the energy levels of the electron donor and the electron affinity of the anion as summarized by Vogel \cite{Vogel}. The cross section peaks when the afore-mentioned resonance condition $I_{p,\textrm{eff}}=E_A+\Delta E$ is satisfied. In our experiments we focused on O$^-$ for which $E_A=$\SI{1.46}{\eV} and $\Delta E=$ \SI{0.55}{eV}. Thus, the pair production of O$^-$ ions is believed to be in resonance with the 5d states of neutral Cs with $I_{p,\textrm{eff}}$ of \SI{2.08}{\electronvolt} (5d $^2$D$_{5/2}$) and \SI{2.10}{\electronvolt} (5d $^2$D$_{3/2}$) as shown in Fig.~\ref{grotrian}. The corresponding normalized cross section as a function of energy difference between the donor and acceptor states for the reaction Cs$^*$~+~O$\rightarrow$~Cs$^+$~+~O$^-$  was taken from the literature \cite{Vogel} and is projected onto the vertical scale in Fig.~\ref{grotrian}. The ion pair production reaction between Cs and O has been studied earlier by Vora et al. \cite{Vora} reporting that Cs(6p) increases O$^-$ production over ground state Cs at lower collision energies, which is encouraging as the ion pair production cross section from 6p states of Cs is smaller than the expected cross section from the 5d states (see Fig.~\ref{grotrian}).

Oxygen (O$^-$) was chosen for our experiments for a number of practical reasons despite of the large endothermicity of the pair production reaction corresponding to relatively high collision energies with an ideal donor compared to e.g. C$^-$ production \cite{Vogel}. Firstly, oxygen has a high electron affinity and, therefore, O$^-$ beams are relatively easy to produce, which alleviates the experimental effort. Secondly, the 5d donor states of neutral Cs relevant for O$^-$ pair production have longer lifetimes than other excited states of Cs, which presumably maximizes the interaction probability. Finally, the 5d states are accessible from a number of upper electronic states, which allows probing the ion pair production hypothesis with multiple laser wavelengths and population pathways. The energies corresponding to each laser used in the experiments are marked in Fig.~\ref{grotrian} with horizontal lines (the line colors correspond to the wavelenghts of each laser and the line widths to the FWHM of each diode laser). Furthermore, the figure displays the closest excited states of Cs accessible with each laser and the branching ratios of the electronic de-excitations leading to 5d states from these upper states. The relevant transitions were identified using the National Institute of Standards and Technology (NIST) atomic spectra database\cite{NIST}. The relative probabilities of each energetically possible transition from the ground state are discussed in Section~\ref{results} where they are used for the interpretation of the experimental data. The branching ratios listed in Fig.\ref{grotrian} were calculated from the oscillator strengths reported in the literature \cite{Warner} taking into account the known degeneracies of the excited states, and finally normalizing the sum of the resulting Einstein coefficients to unity. It is worth noting that the resulting branching ratios differ from those used by Vogel \cite{Vogel}, which is probably due to discrepancies in reported oscillator strengths of the relevant transitions as outlined in the literature \cite{Vasilyev}. Also, the spontaneous lifetimes\cite{Vogel} of each state in \textmu s units are marked in the figure. From the experimental viewpoint these are more important than the branching ratios in determining the expected population densities of the excited states and, therefore, the interaction probabilities of donor-acceptor pairs.   

The conditions for efficient ion pair production are sufficient flux of photons causing excitation of Cs(5d) and adequate density of both Cs(5d) as well as oxygen atoms interacting with each other. If these conditions are met, the ion pair production can occur anywhere along the path of the oxygen atoms sputtered from the ion source cathode (see Section~\ref{setup}) and interacting with the surface layer of Cs atoms and Cs vapor in the proximity of the cathode surface. It is underlined that the ion pair production is not strictly a surface process per se but involves two unbound atoms.

\section{\label{setup}Experimental setup}

\subsection{\label{SNICS}SNICS ion source}

\begin{figure}[!htb]
\centering
\includegraphics[width=\columnwidth]{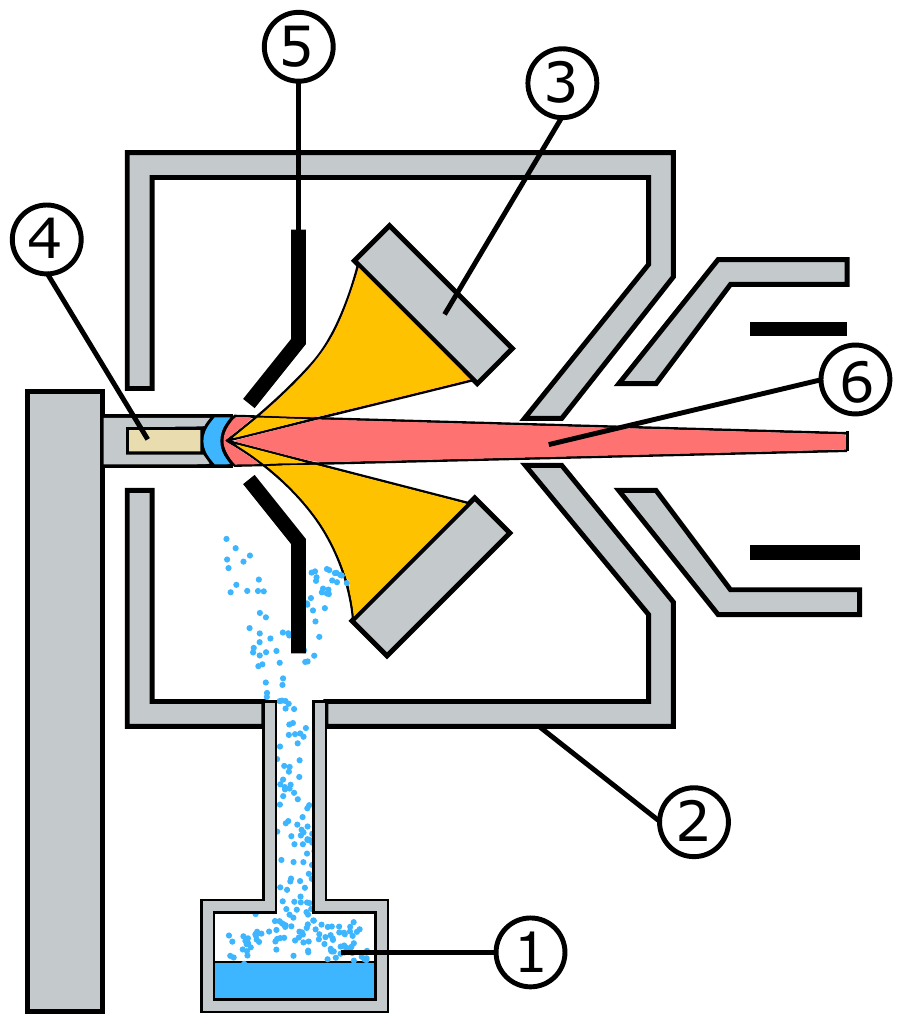}
\caption{\label{SNICS_figure}Schematic drawing of the SNICS ion source. (1) Cesium oven and transfer line, (2) ionization chamber, (3) ionizer, (4) cathode with Al$_2$O$_3$ powder, (5) focusing electrode (immersion lens) and (6) extraction channel and electrodes.}
\end{figure}

The experimental data were taken on a Multi-Cathode Source of Negative Ions by Cesium Sputtering (MC-SNICS) \cite{Middleton_SNICS} by National Electrostatics Corporation (NEC). Figure~\ref{SNICS_figure} shows a schematic drawing of the ion source with most of the detail omitted for simplicity.

\begin{figure*}[!htb]
\centering
\includegraphics[width=\textwidth]{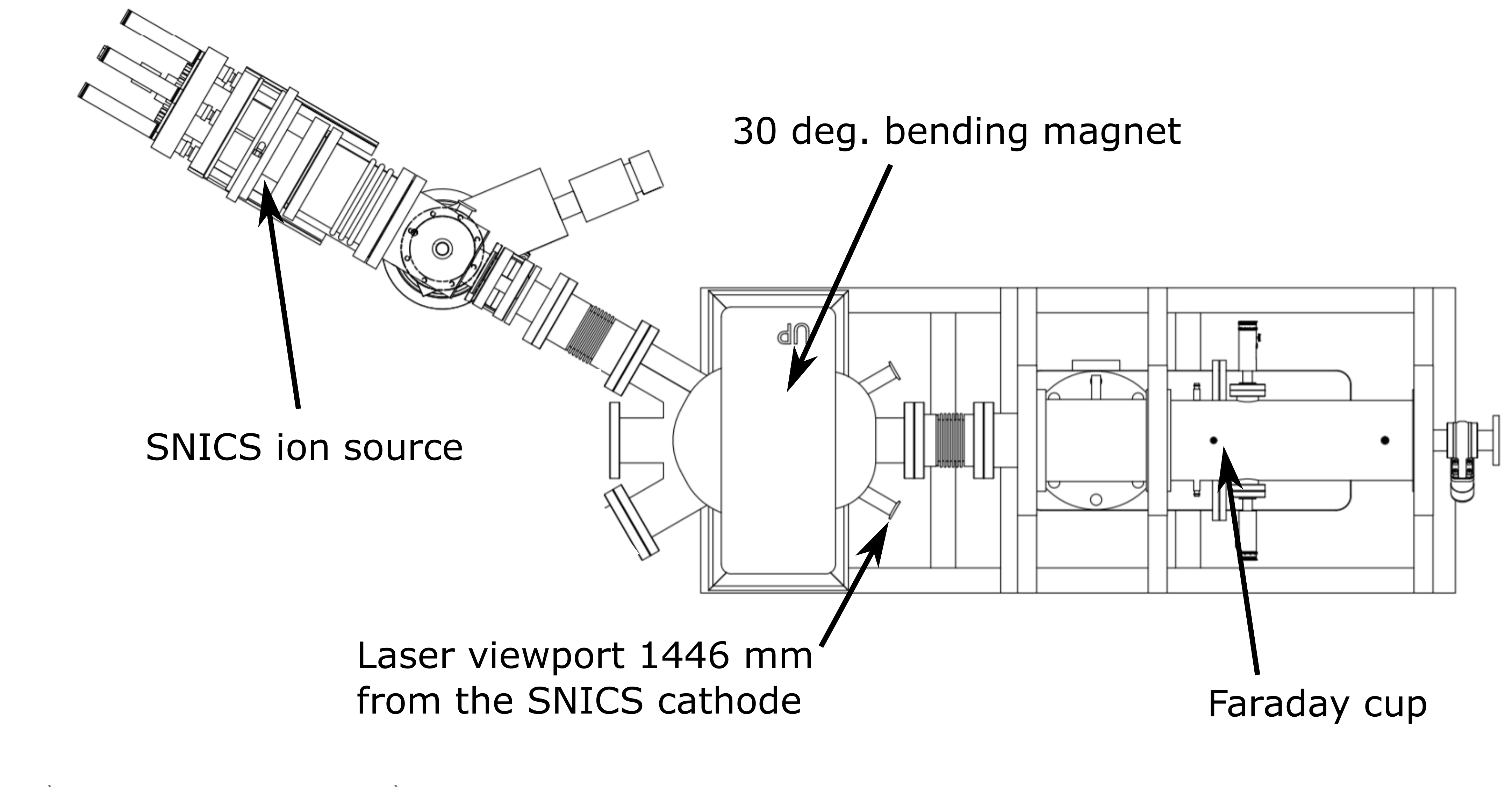}
\caption{\label{beamline}The layout of the MC-SNICS beamline at JYFL Pelletron facility. The multicathode SNICS ion source is connected to the upper branch of the beamline. Other ion sources (to the left) and the Pelletron accelerator (to the right) are not shown.}
\end{figure*}

Cesium (blue in Fig.~\ref{SNICS_figure}) is evaporated from the oven into the ionization chamber where some of it condenses on the surface of the cathode creating a thin, ideally sub-monolayer coverage (greatly exaggerated in the figure) and, thus, lowers the effective work function of the cathode material. Some of the cesium vapor is surface ionized on the hot surface of the ionizer. The Cs$^+$ ions (orange) are accelerated toward the cathode by applying a kV order of magnitude negative potential, typically \SI{4}{\kilo\volt} with the specific source used in this work, and are focused on the front face of the cathode by the "cesium focus lens" or "immersion lens". Increasing the cathode bias enhances the negative ion current through the sputtering yield at the expense of reduced cathode lifetime and beam current temporal stability. In our experiment the cathode was prepared by pressing Al$_2$O$_3$ powder into a cylindrical \SI{1}{\milli\meter} (in diameter) notch on its surface. The O$^-$ ions (red) liberated from the cathode by cesium sputtering are self-extracted from the ion source through the $\sim$\SI{10}{\milli\meter} (in diameter) extraction channel by the cathode potential and finally accelerated to the desired energy further downstream. The ion beam current is adjusted by the user through cesium oven temperature, ionizer temperature, cathode potential and focusing lens voltage. These affect the neutral cesium flux into the ionization chamber and cathode surface, Cs$^+$ ionization probability and flux onto the cathode, sputtering yield and negative ion escape velocity, and beam optics, respectively. 

The extracted negative ion beam was then focused with an Einzel lens into a 30 degree dipole magnet. The transported, mass-analysed, O$^-$ beam current was then measured with a Faraday cup located downstream from the magnet. The low energy MC-SNICS beamline of the JYFL Pelletron facility is shown in Fig.~\ref{beamline}. The distance between the laser viewport and the SNICS cathode is indicated in the figure. Typical beam currents detected from the Faraday cup range from a few nA to several \textmu A. Hence, the current was measured with a low noise current amplifier (Stanford Research SR570), which affects the temporal resolution of the experiment. The implications of the inevitable $RC$-constant are discussed in Section~\ref{results}.     

\subsection{\label{laser}Setup for photo-assisted negative ion production}

\begin{figure*}[!htb]
\centering
\includegraphics[width=\textwidth]{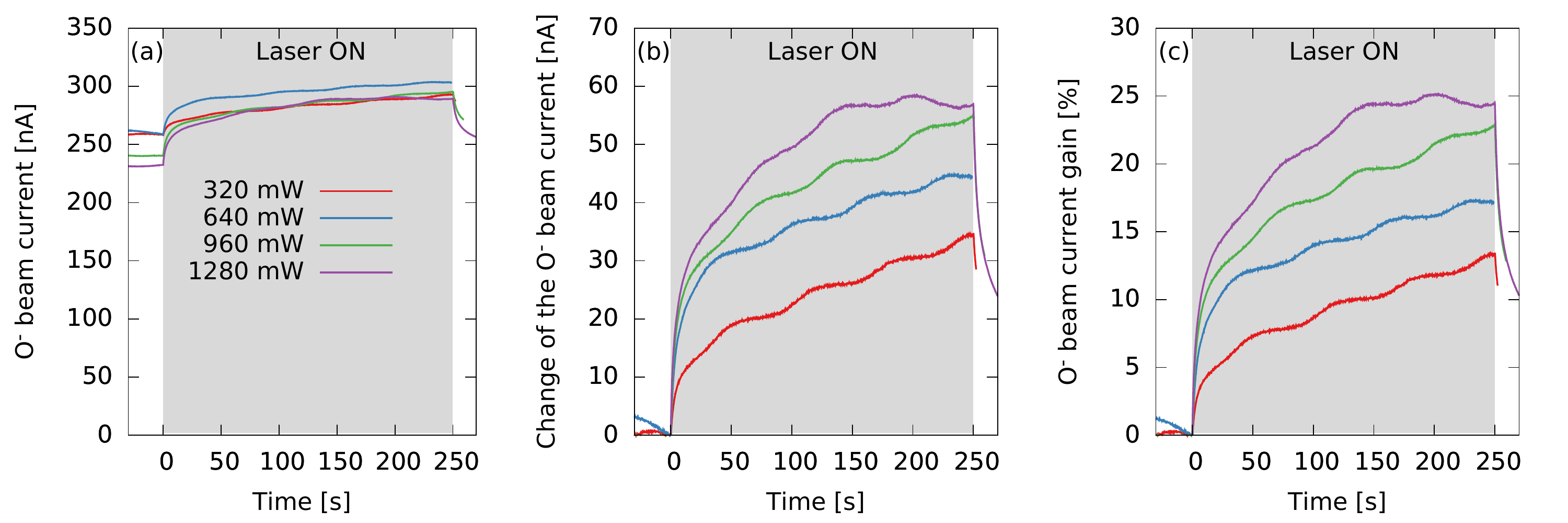}
\caption{\label{powersweep}The effect of the \SI{450}{\nano\meter} laser power on the O$^-$ beam current. The plots show (a) the absolute current, (b) the change of the current in nA and (c) the achieved gain in \% units.}
\end{figure*}

The photo-assisted production of O$^-$ was probed with several diode lasers listed in Table~\ref{lasertable} with their nominal wavelengths ranging from \SI{405}{\nano\meter} to \SI{638}{\nano\meter} and maximum powers from \SI{0.7}{\watt} to \SI{6}{\watt}. These were selected based on the accessible excited states of Cs atoms and expected work function of the cesiated cathode surface allowing to test the pair-production hypothesis. In subsequent figures we have plotted the data using the RGB-colors corresponding to the wavelengths listed in the Table (with the exception of Fig.~\ref{powersweep}).  

\begin{table}[!htb]
\caption{\label{lasertable}The models of the Lasertack GmbH diode lasers, their nominal wavelenghts and maximum powers.}
\begin{ruledtabular}
\begin{tabular}{ccc}
Diode laser & Wavelength [nm] & Power [W]  \\
\hline
LDM-405-1000 & 405 & 1.0 \\
LE-445-6000 & 445 & 6.0 \\
LDM-450-1600 & 450 & 1.6 \\
LDM-520-1000-A & 520 & 1.0 \\
LDM-638-700 & 638 & 0.7 \\
\end{tabular}
\end{ruledtabular}
\end{table}

\begin{figure}[!htb]
\centering
\includegraphics[width=\columnwidth]{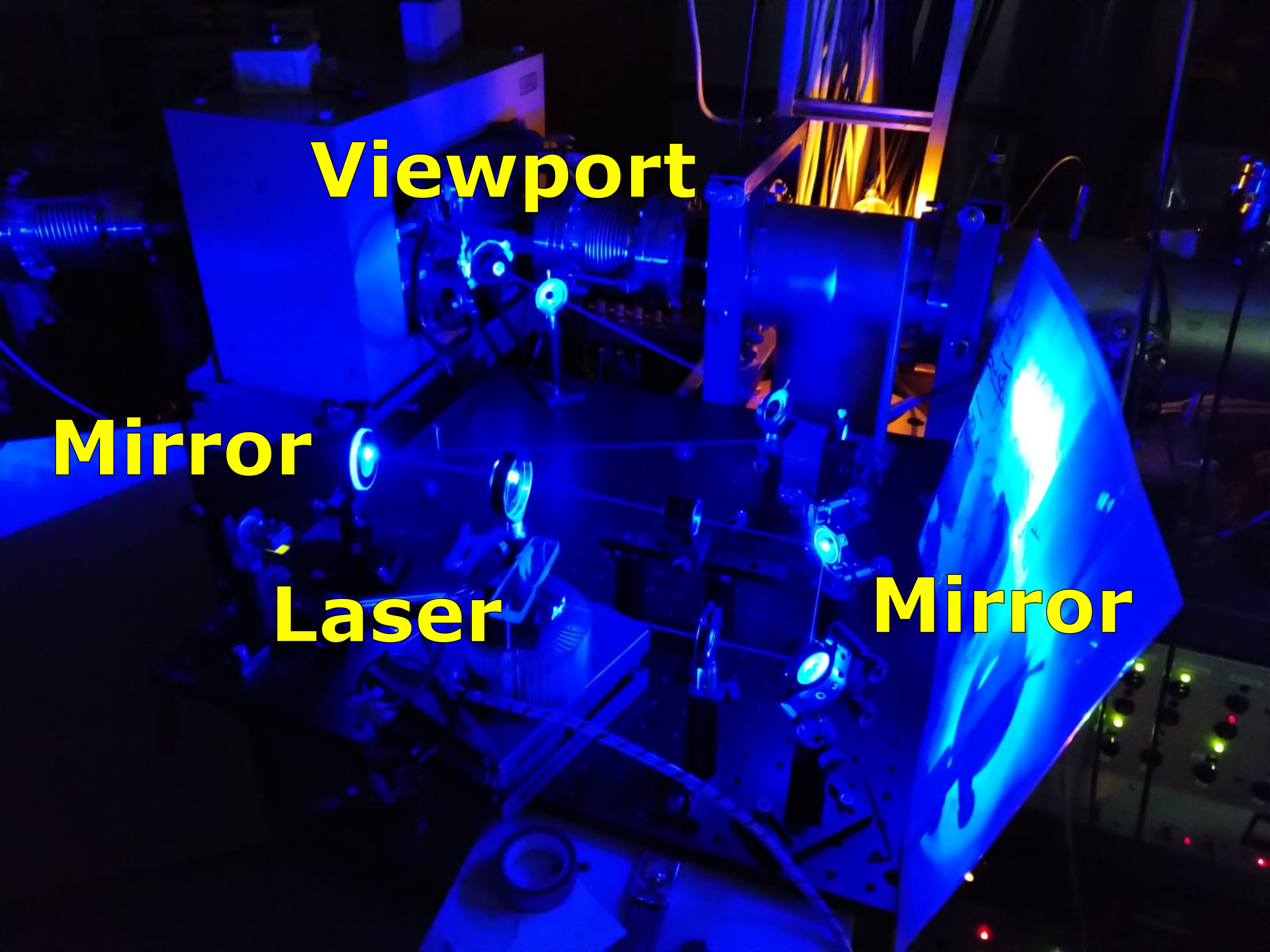}
\caption{\label{laser_figure}The optical table and components used for focusing the laser beam onto the ion source cathode at approximately \SI{1.4}{\meter} distance from the viewport of the bending magnet.}
\end{figure}

The laser beam was focused straight onto the ion source cathode through a viewport of the bending magnet. This arrangement differs from the one used in the earlier experiments\cite{Vogel} where the laser irradiated the volume adjacent to the cathode surface radially. The focal point of our laser setup, shown in operation in Fig.~\ref{laser_figure}, was first adjusted off-line to match the distance to the SNICS cathode, then rotating the laser optics to center the beam spot with the viewport and to illuminate the cathode. The lasers were changed during the experiment using the viewport and two optical apertures shown in Fig.~\ref{laser_figure} as alignment fixtures. The accuracy of this procedure was assessed by measuring the laser power at the SNICS cathode surface by replacing the cathode with a quartz window and using a 1~mm diameter collimator in front of the window, mimicking the cathode cross section. The output power of the LDM-450-1600 diode and the power delivered to the cathode were then measured with a Thorlabs S415C thermal power sensor. It was observed that less than 10\% of the output power (i.e. \SIrange[range-phrase = --, range-units = single]{0.06}{0.09}{\watt} out of \SI{1}{\watt}) reaches the power sensor when the above alignment procedure was applied repeatedly. The maximum power delivered to the cathode in this configuration was less than 20\% (\SI{0.18}{\watt} out of \SI{1}{\watt}) when the alignment was adjusted while observing the power reading "on-line". Altogether, this translates to estimated 5--20\% of the laser power being delivered to the cathode. The power is limited by the mismatch between the beam spot size and the geometrical apertures i.e. extraction channel and sample diameter.

The effect of the laser exposure on the O$^-$ beam current was probed by modulating the output of the lasers by pulsing their drive current (on/off) at various frequencies as well as adjusting the drive signal amplitude to control the laser power. The emission spectrum of the diode lasers depends on their operating temperature and, therefore, the lasers were temperature stabilized to \SI{20}{\celsius} by a Peltier cooling element equipped with a thermocouple feedback control. Nevertheless, it was observed that the emission spectrum shifts slightly with the output power as well as in the beginning of each laser pulse, the emission intensity increasing rapidly with time by $\sim$10\% in the latter case. Both these effects are presumably related to the exact temperature (and its transient) of the light-emitting element, and have implications on the interpretation of the data. Tables~\ref{wavelengthvspower} and \ref{wavelengthvstime} serve to quantify these effects for the \SI{450}{\nano\meter} laser (used here as a representative example) at \SI{20}{\celsius} temperature. The emission spectra of each laser were measured with Ocean Optics USB 2000+ survey spectrometer. Their full width half maximum (FWHM) was measured to be \SIrange[range-phrase = --, range-units = single]{1.4}{2.0}{\nano\meter} depending on the laser power (drive current) and the temperature of the diode.

\begin{table}[!htb]
\caption{\label{wavelengthvspower}The peak wavelength of the \SI{450}{\nano\meter} blue laser with different output powers at \SI{20}{\celsius} set temperature.}
\begin{ruledtabular}
\begin{tabular}{cc}
Output power [mW] & Peak wavelength [nm] \\
\hline
320 & 446.3 \\
640 & 447.0 \\
960 & 448.1 \\
1280 & 449.2 \\
1600 & 449.5 \\
\end{tabular}
\end{ruledtabular}
\end{table}

\begin{table}[!htb]
\caption{\label{wavelengthvstime}The peak wavelength and normalized intensity of the \SI{450}{\nano\meter} blue laser with \SI{1.6}{\watt} final power in different time intervals measured from the leading edge of the laser pulse. The emission spectrum first changes rapidly and saturates in approximately \SI{30}{\milli\second} with the peak wavelength shifting by \SI{1}{\nano\meter}.}
\begin{ruledtabular}
\begin{tabular}{ccc}
Time interval [ms] & Peak wavelength [nm] & Normalized power \\
\hline
0--1 & 448.5 & 0.93 \\
3--4 & 449.2 & 0.98 \\
30--31 & 449.5 & 1.00 \\
\end{tabular}
\end{ruledtabular}
\end{table}

It was later confirmed that the results achieved by pulsing the laser could be reproduced by pulsing the light signal with a mechanical chopper. Pulsing the laser was preferred as it allows studying both, prompt and long-term effects without the added complication of a data acquisition system based on a lock-in-amplifier as described elsewhere \cite{Kronholm}.  

\section{\label{results}Experimental results}

Several experimental campaigns were carried out to demonstrate photo-assisted production of O$^-$ ions and to scrutinize the ion pair production -hypothesis. The experimental results presented hereafter have been organized chronologically in order to allow the reader to follow the reasoning between each step.

\subsection{\label{450nmexperiment}Experiments with the 450 nm laser}

The experiments were started following the footsteps of Vogel\cite{Vogel} i.e. with the \SI{450}{\nano\meter} laser. Absorption of the \SI{2.76}{\electronvolt} photons presumably populates the 7p-states of neutral Cs, which then populate the long-lived 5d-states that are expected to contribute to O$^-$ pair production (see Fig.~\ref{grotrian}). Figures~\ref{powersweep}(a)-(c) show the extracted O$^-$ beam current as a function of time when the Al$_2$O$_3$-cathode was exposed to long (\SI{250}{\second}) laser pulses with different powers. Three plots are shown to account for the inherent temporal variation of the beam current characteristic to the SNICS ion source, causing the initial beam current to vary between data sets. Figure~\ref{powersweep}(a) shows the effect of the laser on the O$^-$ beam current at different power levels, Fig.~\ref{powersweep}(b) the corresponding change of the beam current in nA and Fig.~\ref{powersweep}(c) the gain achieved with the laser normalized to the O$^-$ beam current just before the leading edge of the laser pulse. The contribution of the laser on the O$^-$ beam current is evident with the magnitude of the effect increasing with the laser power (photon flux) despite of the several nm shift of the emission spectrum with power (see Table~\ref{wavelengthvspower}). The beam current, which was constant before the laser was switched on, increases for the whole duration of the \SI{250}{\second} laser pulse implying that the expected prompt effect, namely populating the relevant excited states of Cs followed by ion pair production, cannot alone explain the observed gain in O$^-$ yield.

\begin{figure}[!htb]
	\centering
	\includegraphics*[width=\columnwidth]{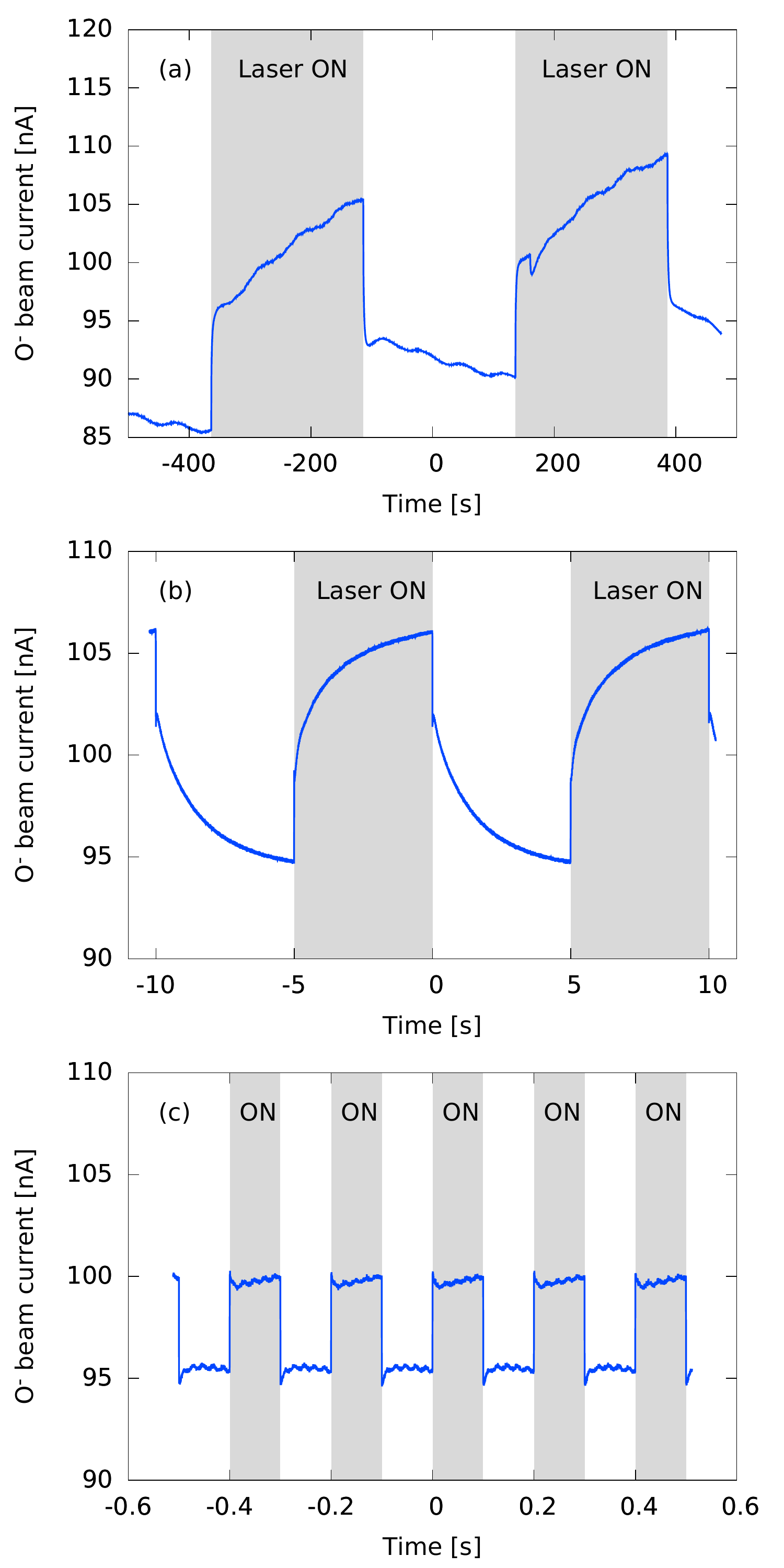}
	\caption{\label{timescales}The effect of the \SI{450}{\nano\meter} laser on the extracted O$^-$ beam current at \SI{1}{\watt} laser power with pulse lengths of (a) \SI{250}{\second}, (b) \SI{5}{\second} and (c) \SI{100}{\milli\second}.}
\end{figure}

The data recorded at varying laser pulse lengths and shown in Figs.~\ref{timescales}(a)-(c) reveal three different time scales in the O$^-$ beam current response to the \SI{450}{\nano\meter} / \SI{1}{\watt} laser exposure. Three different time scales can be clearly distinguished: (a) long-term linear increase lasting for several minutes until the end of the long laser pulses, (b) a logarithmic rise in \SIrange[range-phrase = --, range-units = single]{3}{5}{\second} and (c) a prompt effect when the laser pulse is applied. The first two trends are mirrored (qualitatively) between the laser pulses and the last one at the trailing edge of the pulse.

\begin{figure}[!htb]
\centering
\includegraphics[width=\columnwidth]{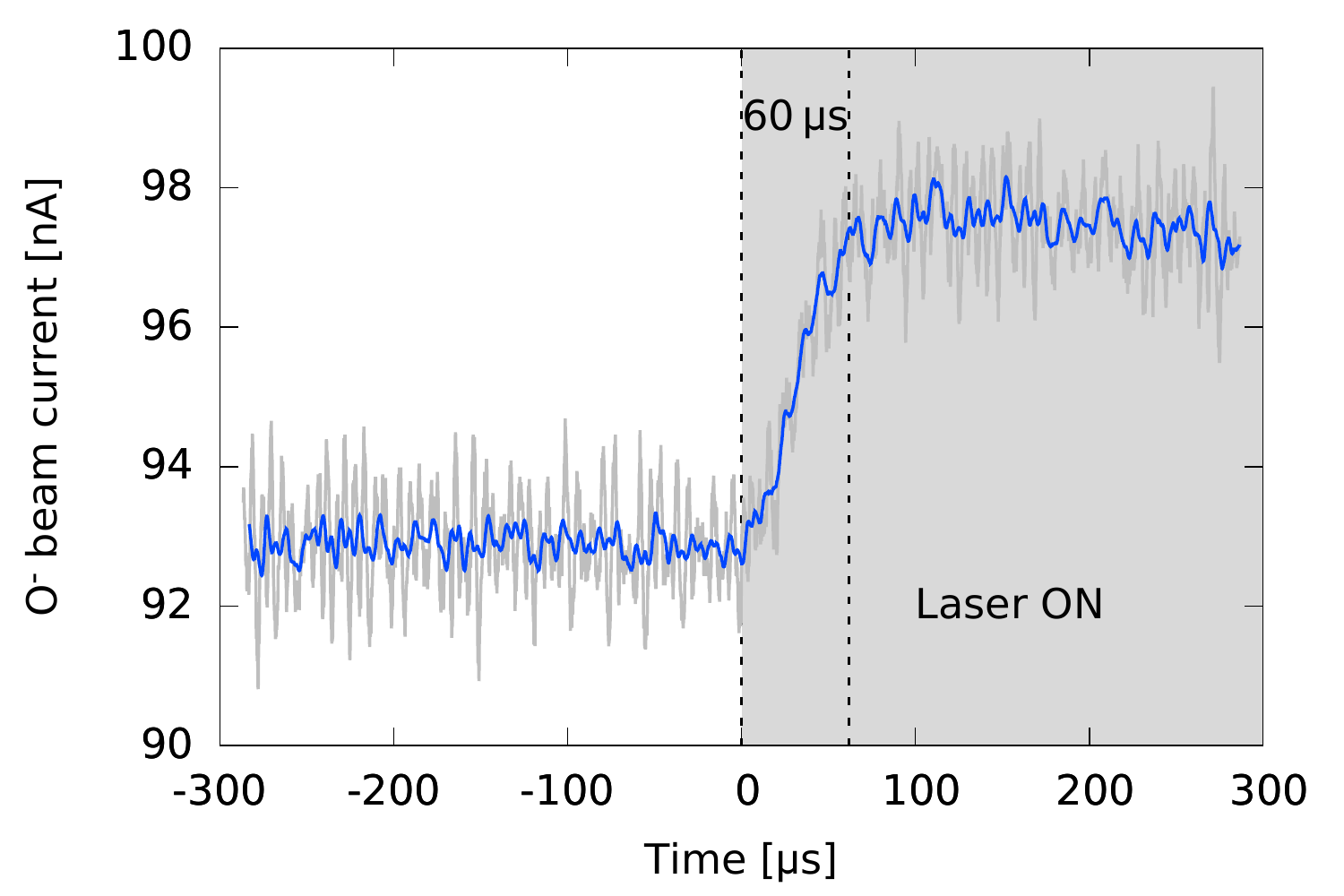}
\caption{\label{risetime}The measured O$^-$ beam current response (raw and smoothed data) at the leading edge of the \SI{450}{\nano\meter} laser pulse. The expected signal rise time corresponding to a step change of current and taking into account the time constant of the measurement setup is \SI{60}{\micro\second}.}
\end{figure}

It was confirmed that the observed prompt effect is indeed instantaneous by measuring the signal rise times at different transimpedance amplifier gain settings and comparing them to the theoretical rise times of a forced step change in beam current. The rise time is determined by the amplifier impedance together with the Faraday cup (approx. \SI{70}{\pico\farad}) and cable (approx. \SI{170}{\pico\farad} at high frequency) capacitances. Figure~\ref{risetime} shows an example of the O$^-$ beam current response at the leading edge of the laser pulse. The expected signal rise time of \SI{60}{\micro\second} corresponding to the calculated time constant of the measurement setup is marked in the figure. It matches with the observed rise time implying that the laser-induced contribution of O$^-$ yield is truly a prompt one. It is important to note that we are not claiming this to be evidence for the ion pair production mechanism but instead argue that the observation confirms the existence of a photo-assisted negative ion production channel (of yet unknown origin).  

The other two time scales cannot be explained easily. The shape of the \SIrange[range-phrase = --, range-units = single]{3}{5}{\second} pulse response is typical to thermal transients that could affect the Cs coverage of the cathode surface by changing the equilibrium density of Cs atoms determined by deposition, evaporation and sputtering. The long-term increase of the O$^-$ beam current could be related to the photon absorption affecting the surface properties, most importantly its work function via a gradual change of the Cs density on the cathode surface. It is worth noting that the ionizer temperature (power) was kept lower than its nominal operational value, which results in modest extracted current. This was done in order to observe the photo-assisted effect superimposed on the continuous beam current signal. The ionizer temperature affects the Cs coverage of the cathode surface by limiting the incident Cs$^+$ flux. Altogether this implies that at constant Cs oven temperature the duration of the long-term transient is sensitive to the ionizer temperature and it varies with the extracted beam current as demonstrated in subsequent sections.    

\subsection{\label{lasercomparison}The effect of laser wavelength - experiments with 450, 520 and 638 nm lasers}

\begin{figure}[!htb]
\centering
\includegraphics[width=\columnwidth]{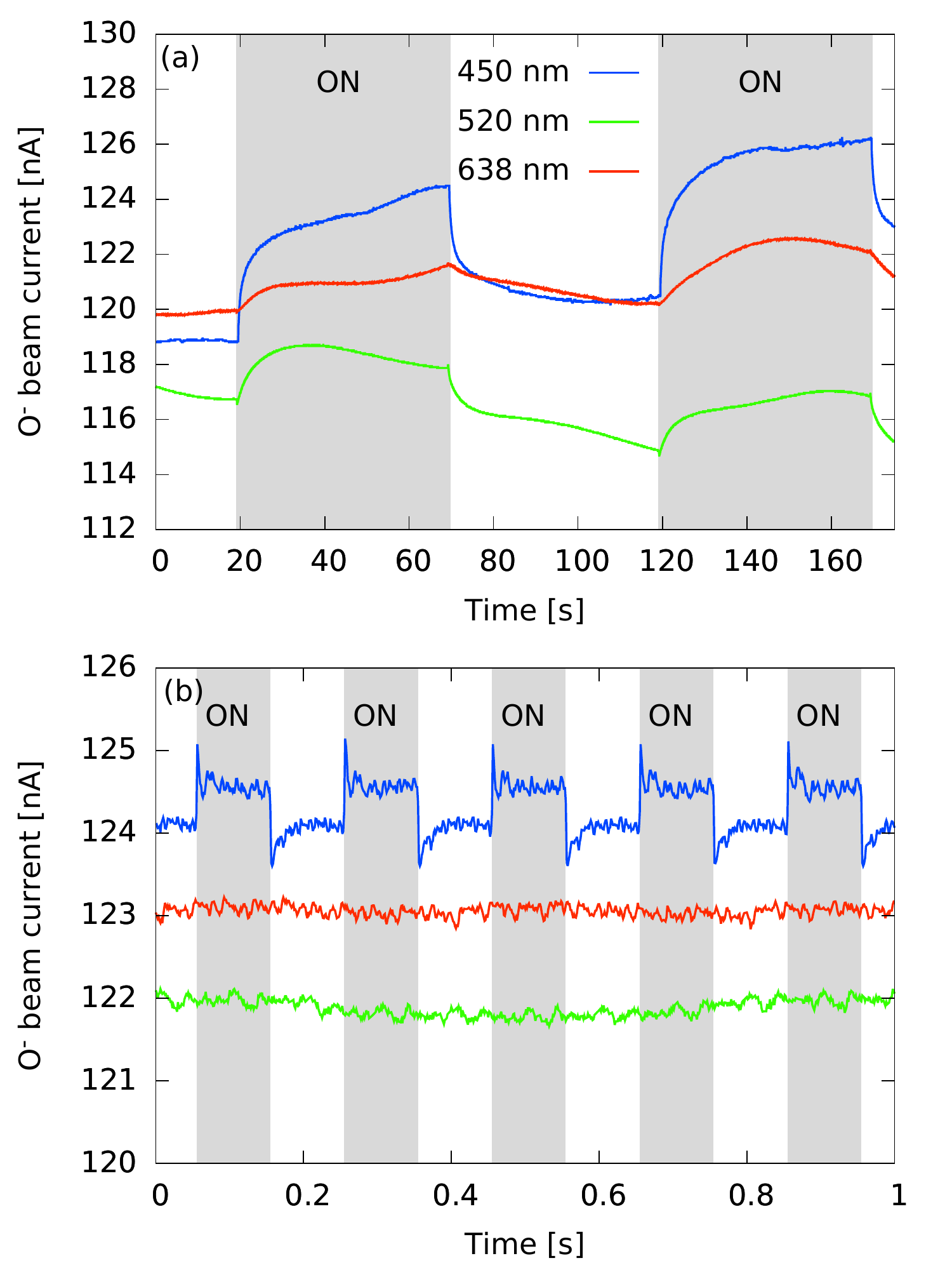}
\caption{\label{RGB}The effect of (a) \SI{50}{\second} and (b) \SI{100}{\milli\second} laser pulses at \SI{450}{\nano\meter} / \SI{1.6}{\watt}, \SI{520}{\nano\meter} / \SI{1.0}{\watt} and \SI{638}{\nano\meter} / \SI{0.7}{\watt} wavelenghts / powers on the O$^-$ beam current}
\end{figure}

In order to establish whether the laser-induced increase of the extracted O$^-$ current is wavelength specific, we first irradiated the Al$_2$O$_3$ cathode with 450, 520 and \SI{638}{\nano\meter} (blue, green and red) lasers. Figures~\ref{RGB}(a)-(b) show the response of the O$^-$ beam current to the above said blue, green and red laser exposures using the maximum power of each diode (see Table~\ref{lasertable}). The long-term effect is observed with all of the three lasers irrespective of their wavelength whereas the prompt effect is induced only by the \SI{450}{\nano\meter} laser, not with the 520 and \SI{638}{\nano\meter} ones. The relative magnitude of the  long-term increase of the O$^-$ beam current matches the difference in total powers between the above said laser diodes, which indicates a thermal origin although the exact mechanism acting on the cathode surface and affecting the beam current remains elusive.

The excited states of Cs that are accessible and most likely to be populated from the 6s ground state of neutral Cs with the \SI{520}{\nano\meter} laser do not populate the 5d states, which are most relevant for the ion pair production mechanism as described above. The 6p states that overlap with the tail of the cross section curve are accessible from the 7s state. However, the 6s $\rightarrow$ 7s transition is optically forbidden, which implies that accessing the 7s state from the ground state would involve a two-photon excitation. Hence, the probability of populating the 7s state with the \SI{520}{\nano\meter} laser can be considered negligible. The \SI{638}{\nano\meter} laser could presumably promote electrons directly to the 5d states but that can be argued to be unlikely as the quadrupole transition strength from the ground state to the 5d states is extremely low \cite{Pucher}. Furthermore, the $>$~\SI{0.1}{\electronvolt} difference of the \SIrange[range-phrase = --, range-units = single]{1.80}{1.81}{\electronvolt} excitation energy and \SI{1.94}{\electronvolt} photon energy is too large for a resonant excitation mechanism. These arguments are in line with the experimental result, namely the the lack of the prompt effect with the \SI{520}{\nano\meter} and \SI{638}{\nano\meter} lasers. 

Similarly, the energy difference between \SI{450}{\nano\meter} laser and the 6s $\rightarrow$ 7p transition could be argued to question the resonant nature of the prompt effect observed with the blue laser presumably populating the 5d states of neutral Cs via excitation to 7p states. This is because less than a 10$^{-7}$th fraction of the \SI{1.6}{\watt} laser power of the \SI{450}{\nano\meter} diode is emitted at the Doppler broadened wavelength of the \SI{455.7}{\nano\meter} 6s $\rightarrow$ 7p$_{1/2}$ transition of neutral Cs. This discrepancy motivated us to continue the experiments with laser wavelengths shorter than \SI{450}{\nano\meter} as explained hereafter.

\subsection{\label{405nmexperiment}Experiments with the 405 nm laser}

\begin{figure}[!htb]
\centering
\includegraphics[width=\columnwidth]{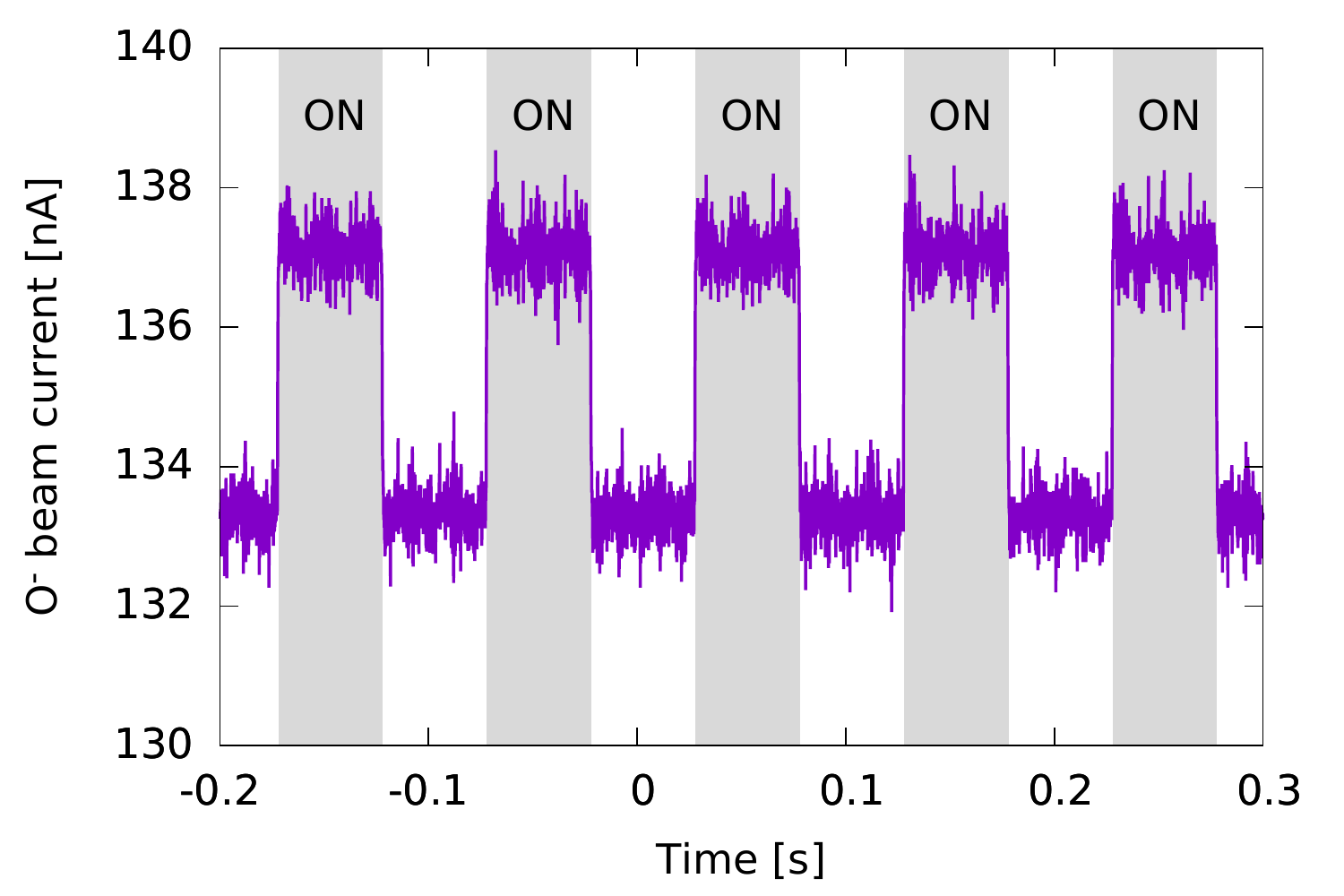}
\caption{\label{405_prompt}The effect of the \SI{405}{\nano\meter} / \SI{1}{\watt} laser exposure on the O$^-$ current from the Al$_2$O$_3$-cathode at \SI{10}{\hertz} laser pulse repetition rate.}
\end{figure}

The experiments were continued with the \SI{405}{\nano\meter} laser, which in principle allows accessing the 4f states of neutral Cs from the 6s ground state and further populating the 5d states (see Fig.~\ref{grotrian}). However, the probability of the 6s $\rightarrow$ 4f excitation is extremely low due to corresponding change in orbital angular momentum being high ($\Delta l = +3$). The prompt effect of the laser exposure on the extracted O$^-$ current with \SI{1}{\watt} laser power is shown in Fig.~\ref{405_prompt}. The absolute effect (in nA) of the \SI{405}{\nano\meter} laser is virtually identical to the effect of the \SI{450}{\nano\meter} laser at corresponding power and O$^-$ beam current. This is strong evidence against a resonant ion pair production explaining the observed prompt effect. This is due to the difference in expected excitation probabilities between the 7p and 4f states from the 6s ground state and subsequent branching to the metastable 5d state altogether suggesting that the 7p excitation should be more efficient catalyst of resonant ion pair production.

\begin{figure}[!htb]
\centering
\includegraphics[width=\columnwidth]{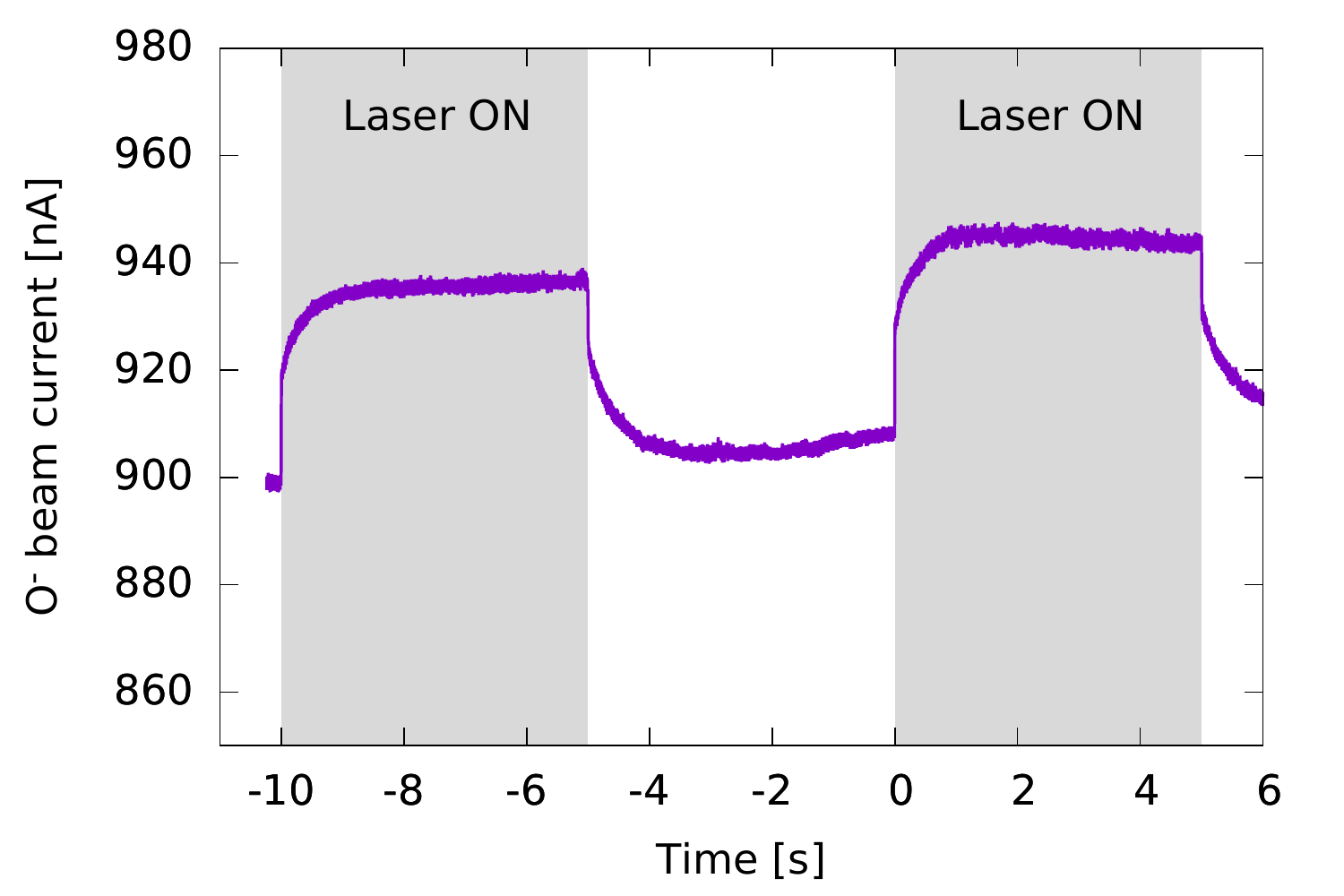}
\caption{\label{405_transient}The effect of the \SI{405}{\nano\meter} laser exposure on the O$^-$ current from the Al$_2$O$_3$-cathode at elevated beam current.}
\end{figure}

Following the discovery of the prompt effect with the \SI{405}{\nano\meter} laser it was confirmed that the photo-assisted contribution on negative ion production can be observed irrespective of the beam current -- an important step in assessing the practicality of the method. The beam current was adjusted by varying the ionizer temperature, i.e. the flux of Cs$^+$ ions impinging on the cathode surface. Figure~\ref{405_transient} shows the O$^-$ current response exhibiting both, the prompt effect and a slow, few second transient.

\subsection{\label{highpowerexperiments}Experiments with the high power 445 nm laser}

\begin{figure}[!htb]
\centering
\includegraphics[width=\columnwidth]{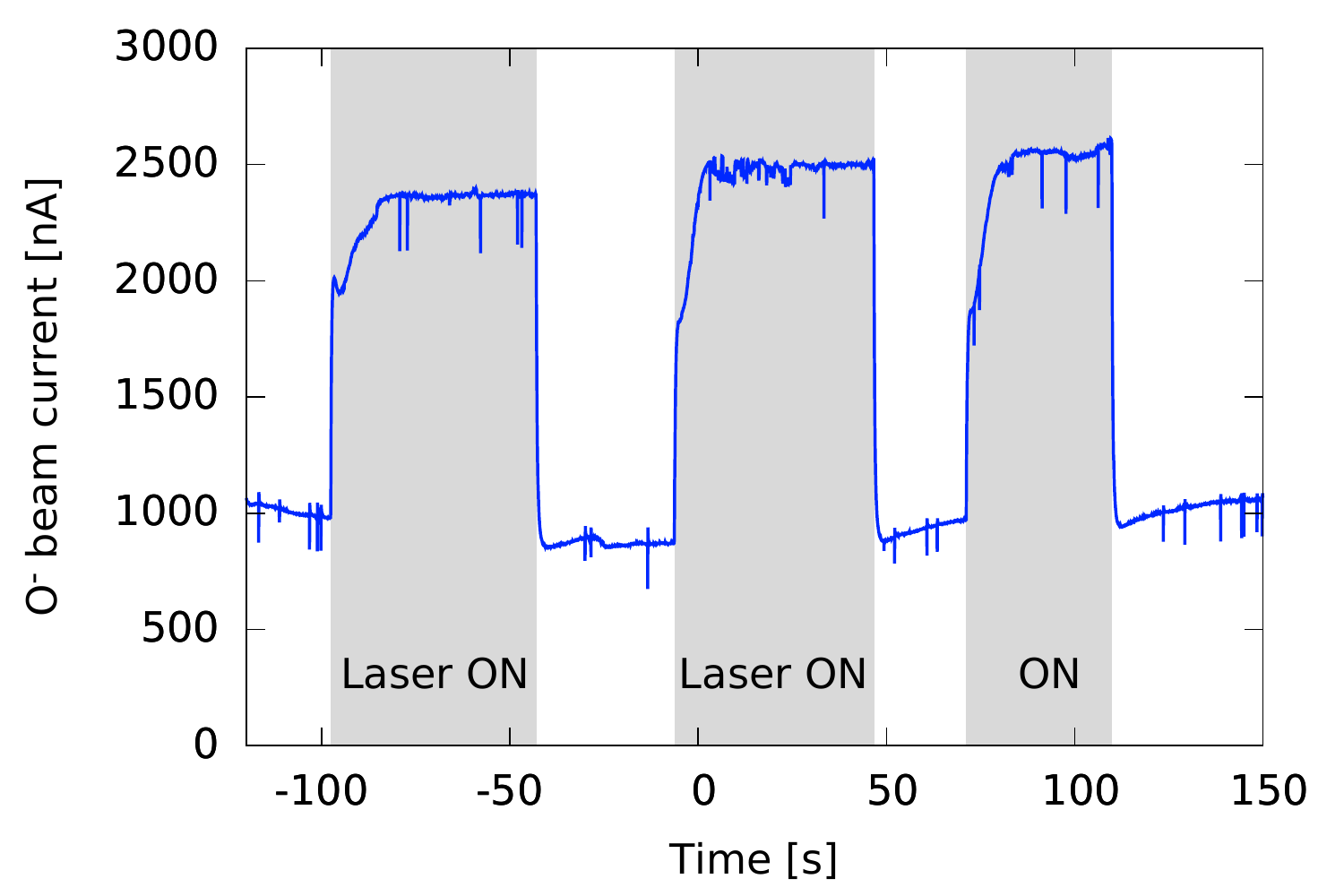}
\caption{\label{445laser_best}The best recorded example of the effect of the \SI{445}{\nano\meter}, \SI{6}{\watt} laser on the O$^-$ current.}
\end{figure}

The data presented in Sections~\ref{450nmexperiment}-\ref{405nmexperiment} motivated us to conduct further experiments with a high-power, i.e. \SI{6}{\watt}, \SI{445}{\nano\meter} laser. The purpose of this campaign was to study the effect of the laser power and wavelength at elevated beam currents as well as observe long-term transients with significant localised power deposition presumably having a prononunced effect on the cathode surface Cs balance. Fig.~\ref{445laser_best} shows the best result obtained with the \SI{6}{\watt} laser at elevated O$^-$ current. Both, the prompt effect inducing a 50--100\% step and the subsequent long-term increase (up to another 35--40\%) of the beam current were observed.  

\begin{figure}[!htb]
	\centering
	\includegraphics*[width=\columnwidth]{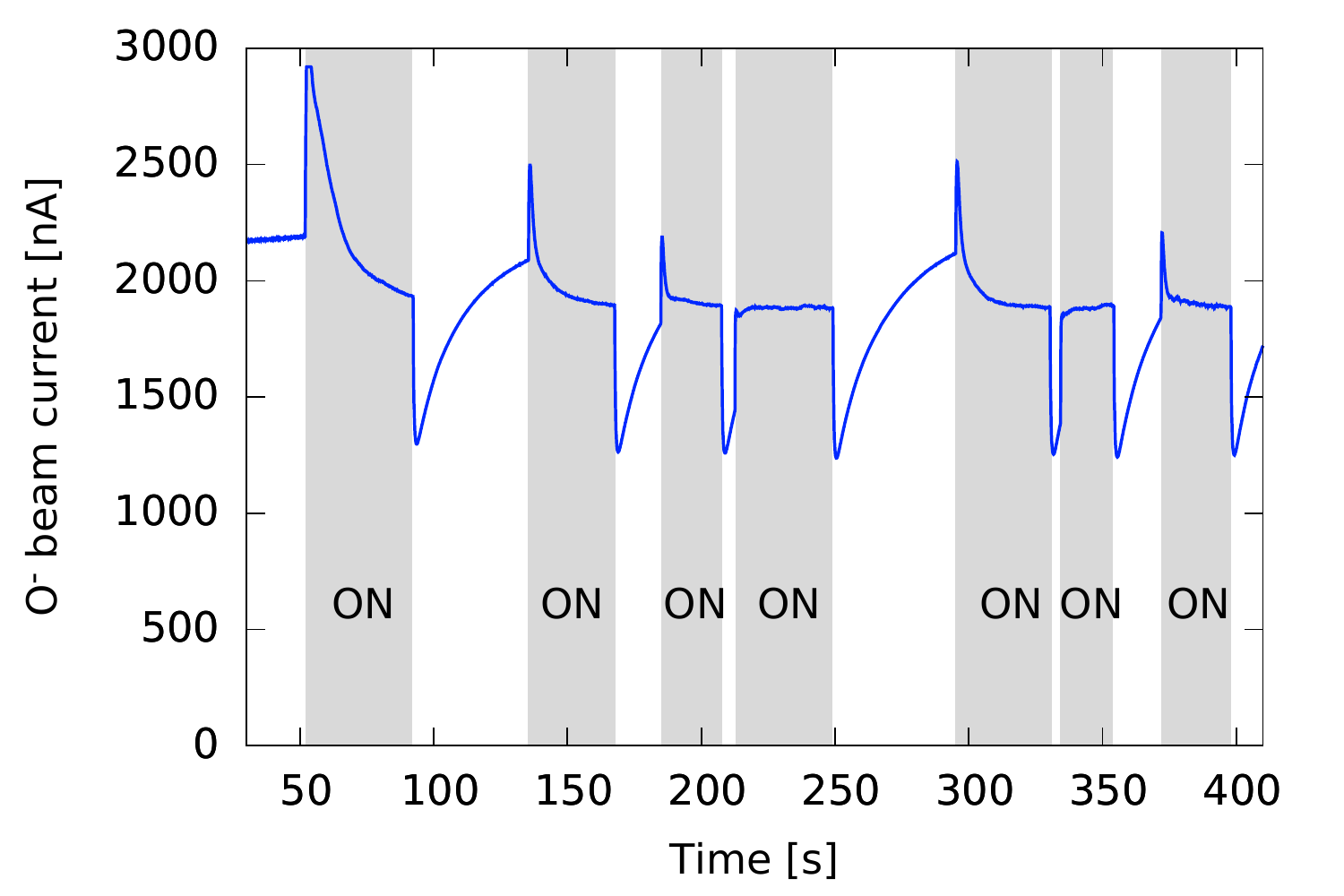}
	\caption{\label{oxygen_445nm_6W_Cs_depletion_long}An example of the effect of the \SI{445}{\nano\meter}, \SI{6}{\watt} laser on the O$^-$ current with randomly varied laser pulse duration and repetition rate.}
\end{figure}

The fact that the \SI{445}{\nano\meter} laser, which should not be efficient in exciting those states of neutral Cs that populate the 5d states, casts doubt onto the ion pair production hypothesis. Nevertheless, the observed photo-assisted negative ion production effect is encouraging as especially the high power laser offers a route to boost the performance of the SNICS source. However, the magnitude and persistence of the effect was observed to depend on the ion source settings. Under some operating conditions the effect fades away i.e. the beam current starts to decrease gradually following the initial increase (prompt effect and gradual rise). It is believed that such long-term trends are due to evolving Cs-coverage of the cathode surface. This view is supported by the data shown in Fig.~\ref{oxygen_445nm_6W_Cs_depletion_long} demonstrating that the magnitude of the prompt effect and the time constant of the gradual decrease of the beam current depend strongly on the time between the laser pulses. When the laser is off Cs presumably accumulates on the cathode surface and is then removed by ablation or evaporation when the laser pulse is applied. In the particular case of Fig.~\ref{oxygen_445nm_6W_Cs_depletion_long} the saturation O$^-$ beam current at the end of the laser pulses is lower than without the laser unlike in the example in Fig.~\ref{445laser_best}. The obvious implication is that the neutral Cs flux from the oven should be adjusted for each set of ion source parameters and laser power.

\section{\label{discussion}Discussion}

The experiments described above have confirmed the existence of a photo-assisted enhancement of negative ion production in cesium sputter ion sources. In the case of O$^-$ ions produced from Al$_2$O$_3$-cathode the effect consists of two components; a prompt effect that appears to be insensitive to the laser wavelength above a certain threshold photon energy achieved somewhere between \SI{2.38}{\electronvolt} and \SI{2.76}{\electronvolt} (corresponding to the \SI{520}{\nano\meter} and \SI{450}{\nano\meter} lasers), and a long-term effect which is presumably driven by evolving Cs coverage of the cathode surface. The experimental setup and the obtained results do not allow quantifying the possible contribution of photodetachment (threshold energy of \SI{1.46}{\electronvolt} for a free anion) by the laser-emitted photons on the extracted O$^-$ current.

The insensitivity of the prompt effect on the laser wavelength disputes the hypothesis of resonant ion pair production in interaction between excited states of neutral Cs atoms (donors) and oxygen atoms (acceptors). Taking into account the discrepancy between the laser emission spectrum and excitation wavelength as well as the modest \SI{3}{\second} time resolution in the earlier experiment with C$^-$ ions \cite{Vogel} it is possible that the authors of that paper were observing the long-term effect instead of a prompt increase of the beam current. Furthermore, the photon-to-anion conversion efficiency deduced from the data given in Ref. \cite{Vogel}, namely \SI{20}{\micro\watt} power (fraction of 10$^{-5}$ of the total \SI{2}{\watt} power) at \SI{455.7}{\nano\meter} wavelength inducing a \SI{5}{\micro\ampere} increase of the beam current, is suspiciously high. These numbers translate into conversion efficiency $\eta=\frac{N_{ion}}{N_{photon}}=\frac{I_{ion}}{P_{laser}}E_{photon}$ of 68\% of the 7p excitations potentially resulting in negative ion formation. Here, $I_{ion}$ is the beam current increase, $P_{photon}$ the laser power at the excitation energy and $E_{photon}$ the laser photon energy in eV. Taking into account the branching ratio from the 7p state to the 7s state, which is considered to be the relevant one for C$^-$ ion pair production, would increase the efficiency above unity, thus violating the conservation of energy even without considering the spontaneous lifetime of the excited state and the interaction probability of the two atoms. We therefore suggest that the effect observed in the cited work \cite{Vogel} is probably due to laser-induced variation of the Cs density (on the cathode surface and in the volume in front of the cathode), which could affect the negative ion production in non-linear manner and would be insensitive to the laser wavelength hence making the 10$^{-5}$ absorption factor irrelevant.

It is emphasized that our experiments do not exclude the possibility of secondary electrons promoting neutral Cs to relevant excited states in front of the cathode and thus contributing to the negative ion yield through the pair production mechanism. Instead we retrospectively question the use of diode lasers for studying the putative mechanism as the observed photo-assisted negative ion production appears to be insensitive to the laser wavelength above a certain threshold energy. It is concluded that the contribution of ion pair production on the negative ion currents extracted from cesium sputter ion sources should be confirmed or disputed with an adjustable wavelength laser scanning across the relevant wavelengths corresponding to excitations of neutral Cs. 

It is possible that secondary electrons emitted from the cathode surface promote Cs atoms to the excited states relevant for the ion pair production as suggested in the literature \cite{Vogel_NIBS}. In this case the laser-induced prompt effect would be best explained by photoelectrons ejected from the low work function surface ($\phi <$ \SI{2.38}{\electronvolt}), and contributing to the negative ion yield by direct attachment or by promoting the ion pair production through enhanced electron impact excitation to the relevant excited states of neutral Cs in the close proximity of the cathode surface. Alternatively, the absorption of photons by the electrons within the cathode material band structure could increase their tunneling probability through the surface potential barrier to the affinity state of the anion \cite{Rasser}. The observed threshold behavior together with the photo-assisted gain of the O$^-$ beam current apparently depending on the Cs balance, and therefore the work function, of the cathode surface are consistent with the photoelectron hypothesis.

Yet another possible mechanism that could explain the observation is photoionization from the metastable 5d states. In this case the laser would promote neutral Cs to the metastable state via the excitation to upper states followed by ionization, i.e. Cs + $h\nu \rightarrow$ Cs$^{*}$ / Cs$^{*}$ + $h\nu \rightarrow$ Cs$^{+}$. Alternatively the first step of the process could be facilitated by electron impact excitation, i.e. Cs + e $\rightarrow$ Cs$^{*}$ + e. Such scheme would explain the fact that the photo-assisted effect was observed in the earlier experiment \cite{Vogel} where the laser was not irradiating the cathode surface and thereby releasing secondary electrons but instead exposing the Cs vapor in front of the cathode. The enhanced negative ion yield in this case would be due to increased Cs$^+$ flux to the cathode and the corresponding change of the negative ion sputtering rate and Cs balance on the cathode surface. Although such effect cannot be excluded, we argue that populating the 5d state via photon absorption and subsequent cascading to the metastable state should be a resonant effect involving the same initial step as the putative ion pair production mechanism. This does not apply if the metastable Cs population was produced through electron impact excitation instead. In that case the enhancement of the Cs$^+$ ion flux and the negative ion yield would be a threshold process with a \SI{2.08}{\electronvolt} minimum energy corresponding to the ionization potential of the Cs(5d) atoms and could, therefore, be driven by the \SI{520}{\nano\meter} laser which was not observed in the experiment. Altogether, our data showing that the prompt effect is not sensitive to the laser photon energy above a certain threshold, not achieved with the \SI{520}{\nano\meter} laser, supports the above explanation based on photoelectron emission affecting the negative ion yield through an unknown mechanism.   

The role of Cs coverage is best illustrated by the experiments with the high power laser using varying pulse repetition rates and pulse lengths, demonstrating that with sufficient laser power and inappropriate Cs coverage the long-term effect of the laser can be adverse. It has been observed\cite{Blahins} that operating the SNICS source in pulsed mode can sometimes (under certain operating conditions) lead to enhanced beam currents although systematic trends covering various ion species were not found. It is plausible that, similar to photo-assisted negative ion production with the laser exposure, the performance of the SNICS ion source in pulsed mode is sensitive to variations of the cathode Cs coverage.

The photo-assisted negative ion production could be of practical importance for the operation of cesium sputter ion sources as demonstrated by the factor of $>$2 increase of the O$^-$ current achieved with the \SI{6}{\watt} laser. Alternatively the method could be applied for reducing the erosion rate of the cathode and, thus, increasing its lifetime by enabling to reach the same beam current (as without the laser) at reduced Cs$^+$ flux. A complete assessment of the method's potential requires experiments with other negative ions and cathode materials, i.e. metals and compounds, especially those that typically have low negative ion yields. The role of Cs could be best studied with cathode materials made of Cs compounds, such as CsCl typically used for the production of Cl$^-$ ions. The transient effects of extracted beam current, presumably caused by the fluctuation of the cathode Cs coverage, could be suppressed in the case of cathode materials with intrinsic Cs content.

It is expected that in cesiated plasma ion sources the benefits of exposing the negative ion production surface to a photon flux from an external source are limited as plasmas naturally radiate up to several tens of percent of the discharge power in UV/VUV-range \cite{Komppula} resulting in significant photoelectron emission from cesiated surfaces \cite{Laulainen}. However, if the follow-up experiments on cesium sputter ion sources with an adjustable wavelength laser were to reveal a significant contribution by the resonant pair production effect, experiments on laser-assisted negative ion production in the discharge volume of cesiated plasma ion sources would be justified. 

\begin{acknowledgments}
MN acknowledges financial support from Her Majesty’s Customs \& Revenues (HMCR) Coronavirus Job Retention Scheme.
\end{acknowledgments}

\end{document}